# ADOPTING E-COMMERCE TO USER'S NEEDS

Mohammad Alshehri[1], Hamza Aldabbas[1], James Sawle[1] and Mai Abu Baqar [2]

[1] De Montfort University, Faculty of Technology, Leicester, United Kingdom
`{mohammad, hamza, jsawle}@dmu.ac.uk`
[2] Albalqa' Applied University, Faculty of Engineering, Al-salt, Jordan
`mai@bau.edu.jo`

## ABSTRACT

*The objectives of this paper are to identify and analyse the extent to which the site is fulfilling all the user's requirements and needs. The related works comprise the history of interactive design and the benefits of user-centered development, which is the methodology followed in this survey. Moreover, there is a brief comparison between Waterfall and User-centered methodology in terms of addressing the issues of time saving and addressing fulfilment of users' needs. The data required to conduct this study was acquired using two research methods; the questionnaire and direct user observation, in order to address all the performance related attributes in the usability stage of the evaluation. An evaluation of the website, based on statements of usability goals and criteria, was undertaken in relation to the implementation and testing of the new design. JARIR bookstore website was chosen as a case study in this paper to investigate the usability and interactivity of the website design. The analysis section includes needs, users and tasks and data analysis, whereas the design phase covers the user interface and database design. At the end of this paper, some recommendations are presented regarding JARIR website that can be taken into account when developing the website in the future.*

## 1. INTRODUCTION

The Internet has brought e-commerce to an entirely new level (Corbitt, Thanasankit, & Yi, 2003) by aiding trading, distribution and sales between organisations, consumers and even between consumers. Over the past few years, the world has faced exceptional advancements in the field of digital communication and online transactions making client's decisions much easier, and allowing greater flexibility. Taking into account the huge amount of information that is constantly available and being transferred between Internet users, the issue of websites usability and the customer-computer interaction arises. It is evident that many firms focus on attracting online customers by utilising a number of methods. These methods comprise provision of services and goods purchased on the Internet and saving customers' time and preventing the barriers associated with distance. However, there are many issues relating to computer-human interaction and interfaces with e-commerce websites, which, if thoroughly investigated, would assist in increasing the usability of such websites, reducing the numbers of customers that are driven-away from sites.

## 2. AIMS AND OBJECTIVES

This paper describes the development of a new version of JARIR Bookstore's website as a case study of research; it will be designed to meet all user requirements and will seek to achieve the highest possible level of user satisfaction and website efficiency overcoming the problems with the original website's lack of usability. The main aims and objectives of this paper are:

- Investigating the user needs to enhance the website's usability and interactive design.

- Analyzing user requirements and fulfilment issues, including user profiling and characteristics such as physical limitations, Computer knowledge, language spoken and attitude towards information technology in general. Furthermore, the activation of online shopping and other features will be evaluated.

- Gathering all the data relevant to the website development process and highlighting the weak aspects of the original website and the most important features and services to offer to visitors.

- Redesigning the interaction in the website considering the usability goals applied to develop such websites, which are navigation, attractiveness, ease of use, efficiency and effectiveness.

- Undertaking an evaluation of the website based on a statement of usability goals and criteria. For example, navigation can be measured by number of clicks on forward and backward buttons to complete a task, whereas efficiency can be measured by time spent trying to obtain a required piece of information from the website.

## 3. RELATED WORK

Many studies have been delivered in the field of design to achieve efficient human-computer interaction. Saffer (Saffer, 2006) defined interaction design as "the art of facilitating interaction between humans throughout products and services. It is also, to a lesser extent, about the interaction between humans and those products that have some sort of awareness that is product with a microprocessor that are able to sense and respond to humans". Interaction design is an art rather than a science; assisting organisations in resolving a set of problems that occur under specific conditions. It requires that different methods be applied to the numerous diverse related issues, such as finding out the best way to ship a product online. The main purpose of interaction design is to enhance communication between two or more human beings or a human and an artificial entity (Saffer, 2006). Carrying out user-centered development aims firms by assisting them achieve a high level of user satisfaction through making the system as easy to use as possible. It is widely agreed that customers naturally seek the easiest way to achieve their objectives (Vredenberg, Isensee, & Righi, 2001). It can be said that complexity appears to be the biggest obstacle obstructing the success of electronic commerce, taking into account that some customers avoid using a given website when it is found to be difficult to navigate or to obtain information from. A usability evaluation is valuable in today's world of web-based activities and e-commerce (Vredenberg, Isensee, & Righi, 2001). Since the 1970's, it has been apparent that the most important part of the usability evaluation is that related to user interface design. Aiming to develop an interactive and usable website requires that the effort and attention of the developer is devoted to the end user when analysing their requirements and preferences. Technology use in the 1960's was confined to specialists developing and maintaining their own applications. The technology revolution has however, enhanced the number of ordinary users accessing information on the Internet. In a related study (Nielsen & Loranger, Prioritizing Web Usability, 2006), usability is defined as "a quality attribute relating to how easy something is to use. More specifically, it refers to how quickly people can learn to use something, how efficient they are while using it, how memorable it is, how error-prone it is and how much user like using it". It is a fact that some firms have been able to double their conversion rate, sales made or product information searched by new users, through implementing usability improvement criteria for their websites. After being redesigned to improve usability, one company website studied improved 135 percent.

### 2.1. Arabic usability guidelines

Perihan Elbaz extracted guidelines to suit the requirements of Internet users from an Arabic background. Having applied the User Acceptance Test (UAT), the author was able to address 715 issues regarding usability on a wide range of Saudi private and government websites (Elbaz, 2011). The following suggestions are the conclusion of the Elbaz's study:

- Home page: all the main announcements should be on the home page and should be visually appealing and easy to recognise.
- Page layout: the space between characters should be sufficient to maximise the readability of the text. Moreover, the header should be large in height, constituting a third of the page and containing a variety of flashes and decorations.
- Contents organisation: the related icons and topics should be grouped into drop-down menus or headings. Users are typically unwilling to waste time searching for appropriate icons or buttons.
- Some Arabic names are extremely long. A name like Almutasembellah has 15 characters. This issue should be taken into account when specifying data-type and length in a given database.
- Applying the concepts mentioned above will highly assist in bridging the gap between the user and the system designer. Furthermore, it should result in a marked increase in the percentage of user satisfaction.

## 4. SYSTEM DEVELOPMENT METHODOLOGIES

Waterfall methodology is a commonly used software development methodology, which was developed in the 1950s. The method became popular in the late 1970s after it was included in software engineering and development books. It evolved as a result of the development of SAGE (Semi-Automated Ground Environment) (Algarwal & Tayal, 2007). The model is referred to as a waterfall due to the cascade effect that occurs during system development, as the developer moves from one phase to another, as in the following illustration. On the other hand, a user-centered development methodology is one of the best methodologies presently applied to building interactive systems (McCracken & Wolfe, 2003). The methodology can be used to develop not only Internet websites, but also network games and spreadsheet programs. This developmental approach, which will be adopted in this paper, has many advantages over traditional development methodologies such as the waterfall (McCracken & Wolfe, 2003). It is user-based, not data based and considers the role of the system user in order to meet all their requirements and expectations. Moreover, it is highly iterative containing a high component of testing and revision prior to implementation. As a result, the user is able to predict how the interface design will be used prior to implementation (Carraro, 2011). In order to conduct a perfect interactive design, there are four basic activities that must be involved (Preece, Rogers, & Sharp, Interaction Design. Beyond Human-Computer Interaction, 2002):

- Identification of the user's needs and establishment of requirements.
- Development of alternative designs to meet those requirements.
- Communication and assessment of the interactive version of the design.
- Evaluation of the version built throughout the process.

## 5. DATA COLLECTION AND ANALYSIS

### 5.1. DATA COLLECTION

One of the most effective methods of information gathering is by questionnaire from which we can gain the demographic information and user experience (Preece, Rogers, & Sharp, Interaction Design. Beyond Human-Computer Interaction, 2009). The means of data collection applied in this paper is a questionnaire designed to cover usability goals. Most answers are responses chosen from a predetermined set of alternatives. Preece et al (Preece, Rogers, & Sharp, Interaction Design. Beyond Human-Computer Interaction, 2009) recommended that; "to work best, the questions need to be short and clearly worded". A questionnaire is a good way to collect demographic information about users (Brinck, Gergle, & Wood, 2001); the questionnaire covered various types of people from different age groups, of both genders, with different levels of education, various jobs and computer skills. Some questions are adopted from Questionnaire for User Interaction Satisfaction (QUIS) in order to enhance the quality of the survey (Preece, Rogers, & Sharp, Interaction Design. Beyond Human-Computer Interaction, 2009). The questionnaire was created using Survs.com website, which is a very useful tool for distributing a questionnaire online via the Internet (emails and social media websites) and obtaining results quickly. Having sent the first copy of the questionnaire over the Internet, a participant responded saying that he could not read English. As a consequence, the questionnaire was been modified to be bilingual (English and Arabic) to comprise more participants in the study.

## 5.2. USER AND TASK ANALYSIS

In this work, we investigated the demographic profile of the website's users; such as gender, age and level of education. Additionally, it explored the user experience of the online shopping website and compares these with the completion of certain other tasks on the Internet and on websites similar to the current JARIR website. All the activities and tasks that the user might expect to be able to conduct via the website should be discussed at this stage.

## 5.3. DATA ANALYSIS AND FINDINGS

Out of the 127 participants who filled out the questionnaire there were 72 males and 55 males, of which 39.7% were between 18 and 35, see figure 1 and 2. The main occupation held by the participants was university student, studying for a variety of educational qualifications. 57.5% were Bachelor's degree students, 27.6% were Master's students with the rest being a mixture of high school and PhD candidates. This indicates that nearly 88% of the participants were well educated, removing the possibility that users might be unfamiliar with using computers and the Internet. Regarding the period spent using computers and Internet technology; 111 respondents stated their long experience with technology in excess of 5 years. This clearly shows the user's familiarity with new technologies and human-computer interaction based systems. The study results identified that the most commonly used Internet browsers were Internet Explorer (IE) and Google Chrome, upon which the newly designed website will be tested. Surprisingly, 107 of the participants had only visited the website between one and five times. This emphasised the clients' significant lack of interest in using the current website, due to usability and interactive design issues. Many of the survey participants declared that their first visit had been that made to fill out the questionnaire.

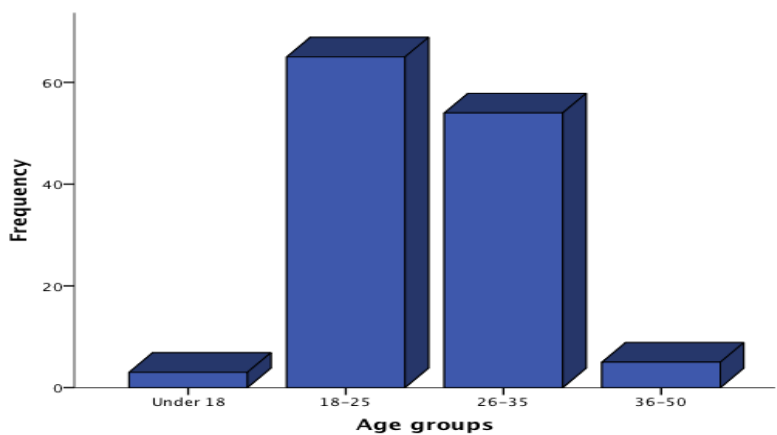

Figure 1. Age distribution of participants

Examination of the table of usability attributes, as shown in Table 1, clearly identifies the main problems that the current JARIR website suffers from. 28% of the participants expressed inconvenience with the website navigation method, claiming that it is misleading and does not assist the user to smoothly navigate the website pages. The user satisfaction with the services provided on the website was attributed with the lowest level amongst all the evaluation criteria. 60 % disliked with the level of services that JARIR presently offers on its website. Just over half of the participants stated that the website is not attractively designed and that it requires some enhancement in terms of interface design and contents. Many visitors declared that there is no appropriate contact offered with the staff, and a lack of a user guide to support online visitors. The website was not found to have any internal errors or unfound pages, but some respondents believed that the website was not functionally correct in the number of pages presented. Moreover, with rates of agreement and disagreement, both around 20%, questions about the ability to learn and remember the website for use on subsequent visits were inconclusive.

| No | Measuring attributes | Level | | | Users' group |
|---|---|---|---|---|---|
| | | Now | Planned | Minimum | |
| 1 | Level of ease of navigation | Low | High | Acceptable | All users |
| 2 | Level of website's content organisation and attractiveness | Low | High | Acceptable | All users |
| 3 | Average time taken to complete a purchase transaction. | N/A | 5 min | 7 min | All users |
| 4 | Number of errors, damaged links and occurrences of not found pages during navigation | 4 | 0 | 0 | All users |
| 5 | Web standards like drop-down menus and proper icon distribution. | N/A | Excellent | Good | All users |
| 6 | Time taken to learn to use website | 15 min | 5 min | 10 min | All users |
| 7 | Average time required to obtain information from the website | 7 min | 2 min | 4 min | All users |
| 8 | The website offers alternative ways for accessibility | No | Yes | Yes | All users |
| 9 | Percentage of navigation simplicity | 33% | 90% | 80% | All users |
| 11 | Percentage of satisfied users | 26% | 90% | 80% | All users |
| 12 | Level of real-time assistance | N/A | High | Acceptable | All users |
| 13 | Users response in-terms of usability goals and satisfaction level | 29% | 90% | 80% | All users |

Table1: Usability Evaluation Table

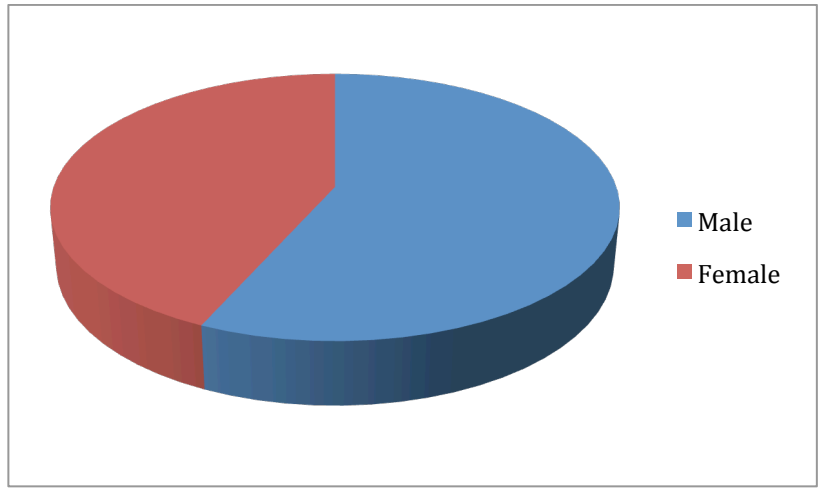

Fig.2. Gender distribution of participants

Approximately 10 of the participants stated that the user's information security was not adequately taken into account when designing the current website. Half of the respondents clearly stated negative impressions of the website whereas only 26 participants owned a good impression of the experience. With shopping being the main intention of the service at the JARIR website, around 90% of the participants had previously visited another online shopping website. 107 out of the 127 participants expressed confidence with shopping online, enhancing the idea that clients value e-commerce services and have no concern with providing bank account details as long as the website visited is deemed to be trustworthy. Implementing an effective e-commerce system appears to be the most often requested feature of respondents. Just over 23% of the respondents wanted online shopping to be one of the futuristic website services made available. Few users raised the need for other languages to be available in addition to English and Arabic. Of the participants, 57 recommended that the implementation of an online shopping feature should include good delivery and coverage map.

### 5.4. CRITICAL CURRENT WEBSITE REVIEW

This section provides a critical evaluation of the current JARIR website, taking into account information gained from the questionnaire results, and the key human factors that have been incorporated into the website design. Having undertaken this step, a number of issues are raised that should be properly addressed as shown in Table 1:

- **Ease of navigation**: The majority of Internet users are extremely concerned about ease of navigation of any website that they visit (Smith & Merchant, 2001). The random distribution of interface contents makes browsing through the website extremely difficult. Without provision of a site map, drop-down menus and headings, finding certain information regarding any product is hard-to-achieve. Navigation tools aid in the creation of mental network maps, which help customers to identify how different sections are inter related. There are various tools that can be employed for creating navigation steps; including drop-down menus, directories, frames buttons, site maps, A to Z index, search engine and colours (Clyde, 2000). Moreover, moving from one link to another in order to complete a task may discourage users from revisiting the website.
- **Design attractiveness**: There is a very famous quote: the 'first impression is the last impression'; thus, choosing appropriate colours and images when designing websites can be critical, due to the different preferences and tastes amongst users. The current main colour and organisation of the website's contents is unacceptable to visitors, giving them a bad impression of the website and also leading to expressions of boredom. Additionally, most advanced websites use graphic designs created using Photoshop or other similar programs in order to attract visitors, which this website does not.

- **Ease of use**: there is confusion regarding duplication of the header and footer on the main page, which makes browsing the website confusing. Another issue is the difficulty experienced when searching for certain items, especially on a first visit; this is due to the site not displaying products in alphabetical order and not providing a search function. The use of terminologies that are familiar to the customer help them to organise an interface and avoid ambiguity (Brinck, Gergle, & Wood, 2001). Additionally, grouping products within an organised menu will lead to more clarity and ease of use when navigating the website. In addition, the size of fonts should be taking into account in case the site's visitors have sight problems; thus the proper size and type of fonts should be considered. Although, some web browsers have the option of controlling the page size (e.g. FireFox and Safari), some customers who have very basic knowledge regarding using computers may not be aware of that option. Providing options for controlling the font size would be valuable and would cover a variety of customers' needs.
- **Functionally Correct**: The number of errors that occur during website surfing should be kept to a minimum. Abnormal disruption to behaviour when visiting a website leads to reduced credibility for customers. The overall impact of these functionality errors is that customers will have doubts regarding the credibility of the website, which may cause them to avoid conducting transactions using the website.
- **Clear and attractive interface**: On an initial visit the interface attraction is low because of silent icons and images. The website also lacks certain high-level design features such as drop-down menus and well-designed components.
- **Learnability**: this measures the time taken to learn how to use the website. Participants should not take too long to learn how to use the website.
- **Efficiency**: The quality and quantity of the information overall is lacking. Most websites nowadays are mainly designed in order to provide the visitor with accurate information (Bhatti, Bouch, & Kuchinsky, 2000). Customers' beliefs regarding the legality and credibility of the website reflect the trust that they place in the website by sharing (or not sharing) their personal and critical information (Wan, 2000). The website does not provide a privacy policy or terms for the user. Consequently, the company may lose its customer's trust when using the website.
- **Accessibility Standards**: An important aspect of the accessibility standards is the foreground and background colour ratio. Special focus should be directed when designing a website on the needs of different categories of customers (Cao, Zhang, & Seydel, 2005). Some people cannot differentiate between certain colours (colour blindness) and as a consequence it becomes difficult for users to clearly see the contents of the website. Similarly some people have disabilities that make it difficult for them to properly use a mouse. For such users it is highly encouraged that the website should be keyboard operable and there should be no keyboard traps present on the website.
- **Easy to remember**: users should be not required to remember information, especially between two pages and two different visits to the website. The current website contents are distributed across pages and are forgettable as there are no hints or images accompanied with the icons (Brinck, Gergle, & Wood, 2001).
- **Security**: One of the most important features of such a website is to allow it to be used to conduct business online. The customer will always be hesitant about online transactions unless a proper method is used for securing them.
- **Satisfaction**: Customer satisfaction maps the number of customers that are comfortable with the design and the methodology that is being used by the website. Customer satisfaction about an e-commerce website is also related to the post delivery business processes including support. Real-time support, security and ease of finding particular items all play an important role in achieving customer satisfaction.
- **Help and assistance**: Since the rise of the Internet, customer services have become an important feature of the online environment. Providing proper support and assistance is critical for managing a successful e-commerce business. Technologies like chat support services, with the help of email, have become the norm in the e-commerce environment. Providing one or all of these services will greatly increase customer's confidence in the website.
- **Overall impression of site**: The website's attractiveness is the backbone of all e-commerce websites. Regardless of ease of use and reliability, if a website is not attractive to customers they will not spend much time on it (Smith & Merchant, 2001). The users' attitude towards the website can only be measured by the customer's response, before and after the introduction of the new system. Customer satisfaction is an important goal, because it is only satisfied customers that spread the word to the wider community and promote the success of the website.

## 6. NEW PROPOSED DESIGN

Having investigated all the issues affecting usability and interaction design in the current website, a new proposal has been developed comprising many user-based modifications to ensure a higher quality interactive interface design.

### 6.1. Header

The new design involves simplification of the logo, in order that clients can easily remember it. The logo is placed on the top left-hand side of the header. As previously explained in usability specifications, some visitors may have medical sight problems, so they can also adjust the font size of the website to make it easier to read; this option is placed on the top right-hand side of the header. The new design comes with two language options: English and Arabic. The current website uses these languages simultaneously which is confusing for visitors. On the new website, the languages appear as flags, so that when the visitor chooses one, the language displayed on the page changes before they begin browsing.

### 6.2. Footer

'Stay Updated' is a feature on the current website which enables customers to remain up to date with regard to new offers provided by the company. It is a good way to encourage customer loyalty to the company and is widely used by competitors in the field. To increase the customer base, 'Find Us' has been implemented as a new way to market products via social media networks and to increase brand strength. Users can visit JARIR pages on Facebook, Twitter and YouTube and remain up to date with the company's most recent offers. 'Locate Us' presents an image integrated with Google maps to help customers quickly find the store address and other contact details. 'Home' navigates back to the home page and 'Feedback' allows customers and visitors to assess the website and give their opinions in order to develop the website via electronic survey. 'About Us and 'Contact' are important to give visitors brief information about the company. Visitors need to be aware of who is running the website and the history of the website as well. Regarding accessibility; this link provides information about the accessibility standards that the website will follow; not only increasing the customer base but also acting to increase the company's credibility as it highlights to follow web standards.

### 6.3. Body

A search box on the right side bar is a tool used by customers to search for any item or product. The search tool has been located in a clear location where most visitors would typically look for a search. Most visitors look for a search box when they first enter a website (Krug, 2009).

Alternatively, an advanced search may be used with a predetermined brand or budget, increasing the efficiency of the website. Nielsen (Nielsen, Search: Visible and Simple, 2001) stated that; "The best designs offer a simple search box on the home page and play down advanced search and scoping". The 'Shopping Cart' is a new tool, which can be used to browse the items added before check out. Before this the customer is also able to select the most appropriate payment method. The customer is also able to provide a discount card, requiring them to give basic feedback concerning the page.

## 7. SYSTEM EVALUATION

- The current system does not execute a real online transaction or payment. This is due to the fact that it requires Merchant account with Master and Visa corporations.
- As this paper is about user interface design, there was not enough time to involve the website security. Data protection on the Internet is a deep topic and needs separate research to integrate the website with up to date information security methods. A technology such as MD5 could be an initial solution for this issue. The security confirmation includes activating Secure Socket Layer protocol in order to protect message transmission online.
- Vischeck.com is a useful tool to check any website accessibility. The website gives a simulation of the website under testing and how visitors suffering from colour blindness would see the interface. Having checked the new website design on vischeck.com, it regrettably gives bad results due to the existence of the colour red. However, JARIR management still insistent on keeping the interface colour as it is presently arguing that it is intrinsic to loyalty and logo value.

## 8. CONCLUSION

It is apparent from this paper that implementation of a human factor based system will assist critically in fulfilling user's needs and expectations. Adopting usability and a more interaction focused design, while developing the system will result in many advantages; such as ease of use, ease of navigation and user satisfaction. Although the waterfall model is a commonly used methodology in the world of system development, user-centered development seems to be a better method when seeking to increase user involvement and time saving. The methods of data collection were the questionnaire and direct observation. Other means such as video typing can be used in future. As clearly stated in the paper, JARIR website lacks many aspects of usability and interaction design. Setting usability goals appropriately assisted in discovering the main problems and addressing them properly. The new database design and system in general will add value to the website once it is adopted by the company's management.

## 9. RECOMMENDATIONS

All of the above human factor recommendations and design considerations are modelled keeping in mind the needs of interactively designed website standards; all of these are critical for reaching and satisfying the community at large. We recommend that all of the considerations should be incorporated into the new website. However, there are some features that have extra costs associated with them. The need for these features and the justification of the costs associated with them are explained below.

1. The shopping cart mechanism requires web payment integration and integration of an inventory management system and is costly to develop. However the incorporation of this feature will enable the website to automate the business process of inventory management, orders, invoicing and billing. The cost of development will pay off with the introduction of the new billing channels and the company will enjoy faster and much more efficient business processes. We highly recommend that this feature should be part of the development list in the future.
 2. We recommend that the management should upload the new website temporarily and request clients to go through it and provide comments on the new design decisions. This will help to understand whether the company meets the needs of the customers with the new website or not. This will result in valuable feedback and will help to further enhance the interface and design of the new website.
 3. Isohunt.com website uses a highly recommended feature, which is voice search. This tool will enhance the accessibility of the JARIR website once it is adopted. It will additionally allow the user to search the website without the need for a keyboard.
 4. Many participants express their desire to have a live chat on the website. This function increases the quality of customer services as well as keeping the customer satisfied with the services provided.
 5. 95% of customers prefer to pay using credit and debit cards. Thus should persuade JARIR management to set up a merchant account with an online transaction company, such as Visa or Mastercard, aimed at enhancing user loyalty and satisfaction.

## ACKNOLEDMENT

Words alone cannot express the appreciation and gratefulness we owe to our parents and siblings. This paper would not have been possible without their unending support and constant encouragement. Many thanks also to JARIR Bookstore management and to all the participants for their assistance with this survey.

**Authors:**

### Mohammad Alshehri

Received his BSc in Management Information Systems from King Abdulaziz University, Jeddah-KSA in 2008 and his MSc in Information Systems Management last year from De Montfort University, Leicester, UK. He is currently planning for resuming his postgraduate study in the field of Human-computer Interaction.

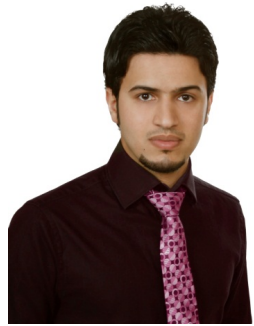

### Hamza Aldabbas

I am currently a PhD student at STRL (Software Technology Research Laboratory), De Montfort University, Leicester - United Kingdom. Obtained M.Sc degree in computer science from Al-Balqa'a Applied University-Jordan (2006—2009). B.Sc. degree in Computer Information System from Al-Balqa'a Applied University- Jordan (2002—2006). I am also a part time lecturer at De Montfort University, involved in teaching and project supervision at B.Sc. & M.Sc levels (2010-until now).

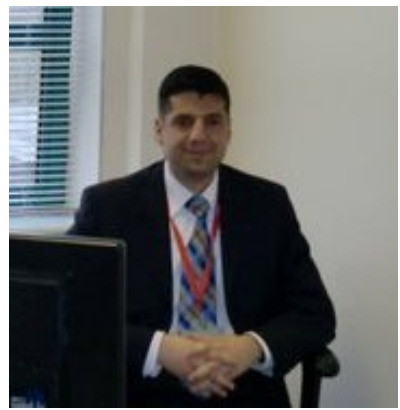

### James Sawle

Received his BSc(Hons) in Computer Science and Mathematics from Durham University in 2009 before completing his MSc in Software Engineering in the STRL at De Montfort University, Leicester. He is currently in the process of a PhD into creativity and search engines. His research interests include IR systems, semantic web, computational creativity and `pataphysics.

### Mai Abu Baqar

Received her B.Sc. degree in Computer Engineering from Al-Balqa'a Applied University-Jordan, before completing her M.Sc degree in Newyork Institute Of Technology (NYIT). Currently she is a lecturer in Al-Balqa'a Applied University/Faculty of Engineering- Jordan.